\documentclass[preprint,journal]{vgtc}            

\onlineid{0}

\vgtccategory{Research}
\usepackage{xcolor}
\usepackage{devins_styles}
\usepackage{enumitem}
\setenumerate{noitemsep,topsep=2pt,parsep=0pt,partopsep=2pt}
\setitemize{noitemsep,topsep=2pt,parsep=0pt,partopsep=2pt}
\usepackage{crossreftools}

\title{

YAC: Bridging Natural Language and Interactive Visual Exploration with Generative AI for Biomedical Data Discovery}

\author{%
  \authororcid{Devin Lange}{0000-0002-3467-0294},
  \authororcid{Shanghua Gao}{0000-0002-7055-2703},
  \authororcid{Pengwei Sui}{0009-0001-0891-0477},
  Priya Misner,
  \authororcid{Astrid van den Brandt}{0000-0002-3676-1341},\\
  \authororcid{Austen Money}{0009-0000-1107-568X},
  \authororcid{Nikolay Akhmetov}{0009-0005-7686-4758},
  \authororcid{Lisa Choy}{0009-0005-1241-6809},
  \authororcid{Marinka Zitnik}{0000-0001-8530-7228},
  and \authororcid{Nils Gehlenborg}{0000-0003-0327-8297}
}
\authorfooter{
  \item
    Devin Lange, Shanghua Gao, Pengwei Sui, Austen Money, Priya Misner,
    Marinka Zitnik, and Nils Gehlenborg are with Harvard Medical School.
  \item
    Devin Lange: E-mail: devin@hms.harvard.edu.
  \item
    Nils Gehlenborg: E-mail: nils@hms.harvard.edu.
}

\abstract{
Incorporating natural language input has the potential to improve the capabilities of biomedical data discovery interfaces. However, user interface elements and visualizations are still powerful tools for interacting with data. In our prototype system, YAC, Yet Another Chatbot, we integrate natural language and interactive visualizations.
YAC uses a tool-calling multi-agent system to generate declarative output, which is interpreted to render linked interactive visualizations and apply data filters.
We also include adjustment widgets, which allow users to directly modify the structured output.
Structured text is also generated to clarify user intent, notify users of system boundaries, and explain aspects of the data with live data element links.
We conducted a user study with domain experts to surface areas where YAC can be improved.
Furthermore we reflect on the capabilities and design of this system with an analysis of its technical dimensions.

}


\teaser{
  \centering
  \includegraphics[width=\textwidth]{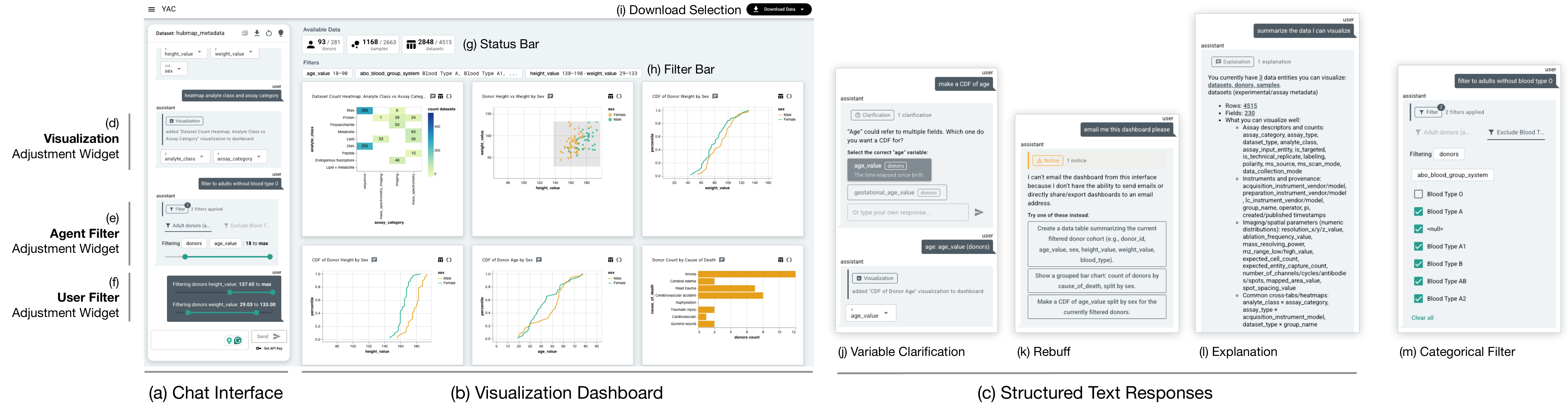}
  \caption{The YAC discovery interface contains two general sections. (a) The chat interface displays user queries and responses from the agent. (b) The linked interactive generated visualizations are displayed in the visualization dashboard, and various structured text responses (c) are displayed in the chat interface. Visualizations can be modified from adjustment widgets (d). Filters can be created from natural language prompts (e) and adjusted with widgets. (f) Selections in visualizations will show adjustment filters in the chat that are linked to the visualization brush. (g) The total number of records in the current filter for each entity is shown in the status bar. (h) The current applied filters are shown in the filter bar. (f) The current selection can be downloaded with the download button. Structured text responses allow the large language model to request clarification (j), redirect impossible requests (k), and explain aspects of the data (l).}
  \label{fig:teaser}
}

\graphicspath{{figs/}{figures/}{pictures/}{images/}{./}} 

\usepackage{tabu}                      
\usepackage{booktabs}                  
\usepackage{lipsum}                    
\usepackage{mwe}                       
\usepackage{ccicons}                   

\usepackage{mathptmx}                  

\begin{document}


\firstsection{Introduction}

\maketitle

Biomedical data discovery is the process by which researchers identify relevant datasets for their scientific inquiry. 
This process can involve searching for experimental results from publications and merging data across those sources. These efforts are time-consuming and require deep expertise to combine datasets appropriately. Thus, biomedical data repositories have been created to perform some of the data-standardization work up front. Such data repositories can make it easier for researchers to find collections of datasets that meet certain criteria. However, even with standardized collections of datasets, finding an appropriate subset can still be challenging. In this paper, we address \textbf{biomedical data discovery} --- the challenge of finding the relevant subset of data within a biomedical data repository.

This process can involve filtering based on metadata attributes of each dataset. Common attributes include the technologies and tools used to derive and process the data from the biological sample, the source organ of the sample, and information about the organism or donor, including demographic, clinical, and genetic information.
Considerable effort has been devoted to building, deploying, and maintaining sophisticated interfaces for navigating biomedical data portals \cite{snyder_human_2019, dekker_4d_2017, roy_elucidating_2023}. However, it is challenging to design an interface that meets the needs of all users who want to consume the data, which can include data scientists, clinicians, and even patients. Even after designers and developers have done their due diligence in identifying user needs, they often find additional requests for specific combinations of visualizations or interactions between them.

One approach in developing and maintaining software is to satisfy the most common use cases very well and not worry about the few edge cases that are not possible with the software. This approach aligns well with van Wijk's proposed model for the value of visualization~\cite{vanwijk_value_2005}, where more users interacting with visualizations for long periods of time is more valuable.
However, in the setting of biomedical research, a key finding that could lead to a cure for cancer may lie in one of these edge cases.
Finding a way to support the wide diversity of edge cases is critically important and cannot be ignored.

This problem touches on a fundamental challenge in designing user interfaces --- there is a tension between the complexity of the interface and its ease of use.
In other words, adding additional features would increase the number of use cases that can technically be accomplished, but also typically increases the difficulty of utilizing those features, sometimes prohibitively so.
What we need is an interface that more flexibly adapts to the users' needs without increasing the burden of interacting with those features.

Natural language interfaces show promise in meeting this need.
Users can often express their goals even when unsure of the required interface actions.
An interface that can interpret these natural language requests and transform the interface for the user has potential.
With the recent advances in large language models, the visualization community is seeing a renaissance of integration of LLMs with visualization systems \cite{pang_understanding_2025, shen_natural_2023, lyi_learnable_2024, liu_ava_2024, wang_data_2024, lee_sportify_2025, zhu-tian_sporthesia_2023, liu_smartboard_2025, shi_nl2color_2024, yan_knownet_2025, vaithilingam_dynavis_2024, kim_phenoflow_2025, dibia_lida_2023}. These systems have explored various ways in which LLMs can enhance visualization systems. However, none of them have explored the problem of biomedical data discovery.

We developed YAC, Yet Another Chatbot, that integrates a large language model with a visualization interface. However, several features make our interface unique.
(1) It uses a tool-based agentic system for generating interactive biomedical metadata visualizations.
(2) It progressively builds multi-view visualizations.
(3) These visualizations are automatically linked through a brushing and linking pattern.
(4). Our system can generate visualizations across multiple related data tables and link filters across those entities.
(5). We bridge natural language interactions with traditional user interfaces for filtering and visualization generation through adjustment widgets.
(6). We include various structured text responses to clarify variables, redirect impossible user requests, and explain with integrated data elements. 
Although some of these ideas have been explored in other systems, to our knowledge, YAC is the first interface to integrate all of these for flexible biomedical data discovery. In addition to designing and developing YAC, we conducted an expert user study and a systematic analysis of its technical dimensions.

\section{Related Work}

\subsection{Natural Language Interfaces}

The promise of natural language interfaces is the ability to capitalize on the expressivity and flexibility of natural language to interact with computers. For interfaces that include visualizations, this extends the goal to interacting with data through natural language and visualizations.
Eviza \cite{setlur_eviza_2016} and DataTone \cite{gao_datatone_2015} are two seminal works that set the stage for an outbreak of NLI research \cite{shen_natural_2023} with many interesting directions.
For instance, resolving ambiguity in natural language queries is an important consideration. Both Eviza and DataTone include ambiguity widgets that enable users to adjust the system's initial response. Setlur and Kumar further explored this challenge by using word co-occurrence and sentiment analysis to map vague requests to specific data modifications and presenting those to users in a way that can be modified \cite{setlur_sentifiers_2020}.
More recently, DynaVis enables users to modify visualizations with natural language requests, while also dynamically generating UI widgets for further adjusting the changes \cite{vaithilingam_dynavis_2024}.
While DynaVis is designed for visualization authoring, we take a similar approach to resolve ambiguous requests for filtering data during a biomedical data discovery session, thereby resolving ambiguity in users' prompts and rectifying mistakes made by the LLM.

Visualization authoring is a common goal when designing NLI systems \cite{wang_natural_2023, shi_nl2color_2024, lyi_learnable_2024, vaithilingam_dynavis_2024}. There is overlap between data analysis systems and visualization authoring systems; however, our focus is on designing a system to support the specific task of data discovery. The visualizations are a means to an end, not the final product to be created and shared with others.
Alternatively, the general task of interpreting a natural language prompt and generating a visualization is a critical component of our system. Cui et al.\ generate infographics from natural language queries about proportions \cite{cui_text-to-viz_2020}. This work focuses on clear and visually appealing displays of relatively simple information, whereas our application needs to generate many connected interactive visualizations. More similar to our use case is NL4DV, which generates Vega-Lite specifications given a dataset and query \cite{narechania_nl4dv_2021}.
Similarly, Luo et al.\ contribute ncNet, a transformer-based model for generating visualization specifications from natural language \cite {luo_natural_2022}.

In addition to generating initial visualizations with natural language queries, NLIs can be designed to support longer interactive sessions that include follow-up queries or modifications of the current application state. Eviza \cite{setlur_eviza_2016} supported such interactive conversations with data, and Evizeon \cite{hoque_applying_2018} went further by applying language pragmatics to characterize the flow of visual analytical conversations. FlowSense \cite{yu_flowsense_2020} integrated natural language queries into an interface for constructing dataflow diagrams. 
Finally, FlowNL \cite{huang_flownl_2023} enables users to update parameters in a fluid flow visualization using natural language.
While these systems share similarities with our prototype, our primary goal is to facilitate data discovery by progressively building a linked, interactive visualization dashboard that can be filtered using both natural language and traditional user interface interactions.

\subsection{Artificial Intelligence for Visualizations}

Artificial intelligence (AI) has been used more broadly than just natural language interfaces \cite{wu_ai4vis_2022}.
For instance, Show Me \cite{mackinlay_show_2007} and Voyager \cite{wongsuphasawat_voyager_2016} produce visualization recommendations given an input dataset using rule-based algorithms, and Data2Vis \cite{dibia_data2vis_2018} accomplishes the same thing with a neural network. However, these do not attempt to capture the users' tasks when exploring the data. GenoREC takes this a step further and incorporates user tasks into the genomics visualization recommendation engine \cite{pandey_genorec_2023}. 

Recent advances in large language models have significantly altered the landscape of HCI Research \cite{pang_understanding_2025}. There have been many visualization systems that incorporate large language models in just the last few years \cite{lyi_learnable_2024, liu_ava_2024, wang_data_2024, lee_sportify_2025, zhu-tian_sporthesia_2023, liu_smartboard_2025, shi_nl2color_2024, yan_knownet_2025, vaithilingam_dynavis_2024, kim_phenoflow_2025, dibia_lida_2023, chen_interchat_2025, wang_data_2025, wen_exploring_2025} and research that studies LLMs \cite{wang_dracogpt_2025, cui_promises_2025, chen_viseval_2025}.
The system that shares the most in common with ours is InterChat \cite{chen_interchat_2025}, which allows users to iteratively update a visualization with linked interactions between the visualization and the chat interface. However, there are a few key differences between our prototype and InterChat. InterChat only displays and updates a single visualization, whereas ours displays multiple visualizations simultaneously. Additionally, we support linked filtering across multiple views through several modalities, interactions with just the visualization, natural language requests, and generated filter widgets. Alternatively, InterChat requires a selection within the visualization and a subsequent natural language request to update the visualization. 
For biomedical data discovery, the ability to quickly and precisely filter the data is an essential task and thus prioritized in our design.

Visualizations with LLMs integrated have been developed for domain-specific tasks before. For instance, KNOWNET \cite{yan_knownet_2025} supports health information seeking, Tailor-Mind \cite{gao_fine-tuned_2025} facilitates self-regulated learning, and several tools have been developed for sports analysis \cite{lee_sportify_2025,liu_smartboard_2025, zhu-tian_sporthesia_2023}. However, there have not been any visualization systems that integrate large language models for the task of biomedical data discovery.

Building domain-specific applications sometimes requires fine-tuning large language models \cite{gao2025txagent,gao_fine-tuned_2025}, and work has been done to create training datasets. For instance, nvBench \cite{luo_synthesizing_2021} is a dataset of natural language queries to visualizations. Similarly, Ko et al.\ provide a framework to generate similar datasets \cite{huang_natural-language-based_2020}.
Although we do not use a finetuned model in YAC, we utilize DQVis \cite{lange_dqvis_2025}, a dataset of natural language queries and visualization specifications about biomedical repository metadata to structure our tool-based approach.

\section{Dataset Description}
\label{sec:dataset-description}

YAC is designed to operate on a single data package. We use the existing Frictionless Data Package standard \footnote{https://datapackage.org/standard/data-package/} which provides a structured way to define multiple entities (tables), their data fields (columns), and the relationships between them.
This package also includes additional information, such as free-text descriptions of entities and fields, the field types (e.g., quantitative, ordinal, nominal), and the range of values a field can have. 

One advantage of using this existing standard is that YAC supports existing biomedical data portals from The Common Fund Data Ecosystem (CFDE \footnote{https://commonfund.nih.gov/dataecosystem}), a major program of the United States National Institutes of Health (NIH) to build infrastructure for data-driven biomedical research. CFDE publishes the metadata for those portals in the Crosscut Metadata Model (C2M2) \cite{charbonneau_making_2022} format --- an extension of the Frictionless Data Package standard.
An important distinction with these data packages is that they represent the metadata of the repositories. These tables often include metadata about the organism (e.g., age, sex), the biological sample (e.g., what organ was the sample collected from), and the dataset (e.g., what technique was used to derive the dataset from the sample). Each row of the dataset metadata file will then contain a field (either an ID or a URL) that directs to the related dataset file (e.g., BigWig, BAM/SAM). While a clear boundary between metadata and data is often hard to define, in our work, all information loaded into database tables is considered metadata, while information that is stored in files on a file system is considered data.

We have tested YAC on four different datasets, two of which will be discussed in detail in this paper. Three datasets are from biomedical data portals of consortia funded by the United States National Institutes of Health (NIH) that are part of the Common Fund Data Ecosystem (CFDE). \textbf{HuBMAP}, from the data portal of the NIH Human BioMolecular Atlas Program~\cite{snyder_human_2019}, contains 320 data fields across three entities, with the largest file containing 4,515 rows and is the dataset used in our user study. \textbf{SenNet}, from the data portal of the NIH Cellular Senescence Network~ \cite{lee_nih_2022}, contains 35 data fields across six entities, and the largest file has 2,740 rows. Finally, \textbf{4DN}, from the NIH 4D Nucleome Data Portal~\cite{dekker_4d_2017, roy_elucidating_2023}, contains 101 fields, 20 entities, with a maximum file size of 50,570 rows.
To illustrate YAC with a simpler, more familiar dataset, we also include the \textbf{Palmer Penguins} dataset \cite{gorman_ecological_2014, horst_palmerpenguins_2020}, which contains a single data table, nine data fields, and 344 rows.

\section{Design Requirements}

The high-level goal of biomedical data discovery is to find a collection of datasets from  a larger set that meet a criteria (both fuzzy and crisp). To guide our development, we created several design requirements.

\begin{enumerate}[align=left]
\setlength\itemsep{0.5em}

\item[{\crtcrossreflabel{\textbf{Filter-Q}}[r:filter-q]}] In some cases, it is necessary to filter the data based on the values of a \textit{quantitative} field. For instance, only include donors of a specific age. Multiple filters should build on each other. E.g., only donors of a specific age and with a height of over 6 feet.

\item[{\crtcrossreflabel{\textbf{Filter-C}}[r:filter-c]}] Similarly, filtering the data based on \textit{categorical} values is essential. For instance, only include biological samples that come from a specific organ.

\item[{\crtcrossreflabel{\textbf{Filter-ER}}[r:filter-er]}] Since the goal is to create a list of datasets, the filtering must be applicable across \textit{entity relationships}. For instance, filtering donors in a specific age range should also filter biological samples from those donors and datasets derived from those samples.

\item[{\crtcrossreflabel{\textbf{Characterize}}[r:characterize]}] After performing a filtering operation, users must be able to characterize the resulting subset of the data. This step is essential for understanding the result of the filter and for reasoning about whether the resulting subset is acceptable.

\item[{\crtcrossreflabel{\textbf{Refine}}[r:refine]}] The process of filtering data and characterizing the result is iterative. Supporting a fast cycle of filtering and viewing the data is essential for efficient refinement of the dataset collection.

\item[{\crtcrossreflabel{\textbf{Complement}}[r:complement]}] Actions to update the application state can occur through the user interfaces or natural language. The action must be possible in the mode where it can be most effectively made so the strengths of each mode can complement the other. Actions that map well to both input modes should be supported in both.

\item[{\crtcrossreflabel{\textbf{Disclose}}[r:disclose]}] Actions that the LLM performs, such as filtering, should not be hidden from the user. It should be clear when and how the application state is updated by transparently communicating it in the interface.

\item[{\crtcrossreflabel{\textbf{Adjust}}[r:adjust]}] The user should be able to adjust the actions that the LLM has taken, either to resolve ambiguity in the user's query, correct a mistake the LLM has made, or adjust the selection quickly after new information leads to an updated understanding.

\end{enumerate}

\section{YAC System Design}

YAC is a system that integrates a chat-based interface (Figure~\ref{fig:teaser}a) with a multi-view visualization (Figure~\ref{fig:teaser}b).
The chat interface allows users to enter natural language queries, which are interpreted by a multi-agent system that returns structured responses through tool calls.

\subsection{Multi-Agent Orchestration}
\label{sec:multi-agent}

\begin{figure*}[tb!]
	\centering
	\includegraphics[width=\linewidth]{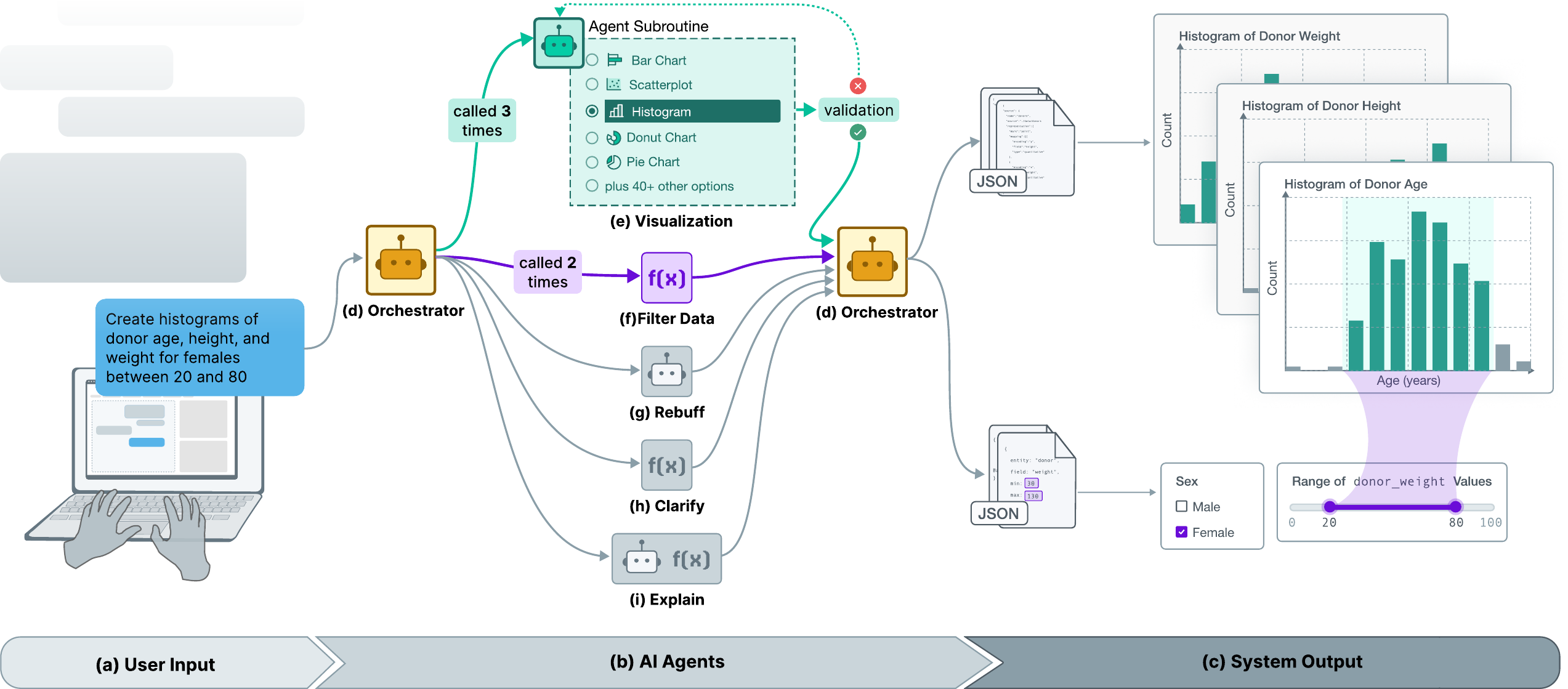}
	\caption{When a user submits a query (a), it is sent to a tool-based multi-agent system (b). This system consists of an orchestrator (d), which will select from available tools In this case (e), a visualization generation tool and (f), a filter tool. Tools consist of a combination of functions and subagents. In this example the structured text based responses (g--i) are not invoked. Once all subroutines are resolved, the orchestrator forwards the structured output that will be interpreted and displayed in the system (c).}
	\label{fig:technical_overview}
\end{figure*}

When a user types a message into the chat interface, it gets processed by an orchestrator agent (see Figure~\ref{fig:technical_overview}b). This orchestrator agent has information about available data provided as context.
Additionally, a list of structured tools is provided. Each tool is defined as a function with a name, description, and parameters. YAC contains five tools: \code{CreateVisualization}, \code{FilterData}, \code{ClarifyVariable}, \code{Explain}, and \code{Rebuff}. The orchestrator selects one or more tools and generates their arguments. For some tool calls, such as \code{CreateVisualization}, the function arguments are passed to an additional agent for further refinement.

The subagents can make subsequent inference calls, but can also include deterministic actions such as a validation step or dynamic data linking. The subagent's final result is passed back to the orchestrator. Once all tool calls are resolved, the full list is passed back to the client for processing. The client will then update the system to render visualizations, apply filters, and display messages in the chat interface.

\subsection{Visualization Grammar}
\label{sec:grammar}

We designed a grammar for biomedical metadata visualizations to use in YAC. Our grammar provides an abstraction layer to produce both visualizations and tabular representations of data, linked through selections and filters.
Our grammar is similar to Vega-Lite \cite{satyanarayan_vega-lite_2017}, and our visualization toolkit renders some specifications by creating Vega-Lite specifications and using the Vega \cite{satyanarayan_reactive_2016} view for more advanced interaction operations.
However, our grammar provides two features that are important for YAC and are not possible with just Vega-Lite. First, our grammar supports tabular representations of data, where each row in the table corresponds to a single row in the input data. 
Tabular representations are ubiquitous in biomedical data portal interfaces, and thus, we include them in our grammar.
The second feature is cross-specification linking of filters and selections, which decouples layout from the grammar and allows visualizations to be arranged freely using standard HTML.

\subsection{Visualization Generation}
\label{sec:tool-based-vis}

In YAC, visualizations are generated through a tool-calling system \cite{gao_democratizing_2025}. Each tool corresponds to a combination of visualization encodings or, more colloquially, chart types. These tools are derived from the template visualizations in the DQVis dataset \cite{lange_dqvis_2025}. Each visualization includes a free text description, example questions the visualization answers, design considerations, and common tasks the visualization is useful for. This information is provided to the LLM as part of the tool description.
Parameter descriptions specify how each variable maps to a visual encoding.
Each template visualization has a corresponding visualization specification in our grammar.
The visualization specification is not exposed to the visualization agent directly; in principle, alternative visualization grammars could replace ours without affecting the agent's behavior.
The template-based approach improves stability, but reduces the overall flexibility for the visualizations the system can generate. Therefore, we also include a general-purpose visualization generation tool, with the full visualization specification exposed as a function parameter.

Since the user can request multiple visualizations from a single query, e.g., ``Show me the distribution of donor age, weight, and height in three histograms,'' and the visualization agent produces only a single visualization, the orchestration agent sends a separate request for each visualization, with context describing what that visualization should show. In this example prompt, three requests will be made, each with a different description of the required visualization. In addition to selecting a visualization tool (e.g., histogram), the visualization agent must select data fields to pass (e.g., age). After visualization generation, a deterministic validation step checks the variable choices. This includes ensuring that the data fields exist in the data schema and checking for other types of mistakes, such as selecting the same data field to encode both X and Y positions. If an error is found, the error message is sent to the visualization agent with a request to fix the issue. 

\subsection{Filter Generation and Selections}
\label{sec:filter-generation}

The orchestration agent can call a data filtering tool. This tool supports two filter types: point and interval selections.
Point selections define a list of valid values for a data field, and interval selections define a numerical range on quantitative data fields.
The frontend system of YAC uses these filter definitions to update the data displayed in the visualization dashboard by combining filters created by the filter agent and the user, and applying filters across related entities.

\textbf{Filtering Logic}
All filters are applied to every visualization and update dynamically as filters change, either through user interaction or as the filter agent updates the state. When multiple filters are present, they are combined so that only records that meet all criteria are included in the visualization. Furthermore, filters across entities are supported. For instance, if a filter is applied to show only male donors, only biological samples from male donors are included in the visualizations. Alternatively, if there is instead a filter applied to the samples entity, such as only showing samples from the liver, then only donors who have donated at least one liver sample would be included.

\textbf{Selection from Visualizations}
YAC supports brushing on visualizations by augmenting the LLM-generated specification with a selection determined with a rule-based system.
The system inspects the field encodings in the specification.
We consider only fields present in the source data, as derived or transformed fields may not support direct filtering. If any quantitative fields are encoded with X or Y positions, YAC includes a 1D interval brush on either the x-axis, y-axis, or a 2D interval brush (\ref{r:filter-q}).
If no quantitative fields qualify, categorical fields encoded with X position, Y position, or color are used instead, with the point selection targeting all such fields jointly.

\subsection{Adjustment Widgets}
\label{sec:adjust-widget}

Adjustment widgets are user interface components that both describe (\ref{r:disclose}) the action taken by the multi-agent system, and support adjustment (\ref{r:adjust}) of those actions. This applies to both visualizations and filters.

\textbf{Visualization}
The visualization adjustment widgets enable users to modify the data fields represented in the visualization. The data fields are identified by checking every field value in the representation layer of the specification. For fields that exist within the source data --- as opposed to fields derived in the data transformation layer --- a dropdown is introduced in the widget. This dropdown lets the user swap out a different field of the same data type.
This approach is distinct from a fully fledged visualization editor that supports arbitrary modifications. 
This widget is designed to support quick and easy adjustments that are often desired.

\textbf{Filter}
The filter widgets allow users to adjust generated filters(Figure~\ref{fig:teaser}e,m).
The user can change the target entity and field, and adjust the range of values for the filter. This ability is essential for multiple reasons.
First, these serve as a fallback guardrail in scenarios where the multi-agent makes mistakes. Second, users may ask an inherently ambiguous question. Our multi-agent system cannot determine what 'old donors' means for each user. But it can make a reasonable selection and allow the user to make a quick adjustment (\ref{r:adjust}). Finally, the user may not yet know what the best filter selection is. After making an initial filter and viewing the updated visualizations, they may want to adjust the filter (\ref{r:characterize}, \ref{r:refine}).
Quickly altering selections ---
performing hundreds of iterations of data characterization and filter refinement within seconds ---
is one way that YAC leverages the strengths of user interfaces over repeated text-based requests (\ref{r:complement}).


To reinforce the connection between agent actions and user actions, we also generate the same filter widgets when the user makes a filter by brushing a visualization (Figure~\ref{fig:teaser}f). This results in a mirrored interaction between the visualization brush and the corresponding filter widget in the chat interface --- modifying either will update both.

\subsection{Structured Text Responses}
\label{sec:structured-text}

YAC supports several structured text response types (Figure~\ref{fig:teaser}c), each implemented as a distinct tool available to the orchestration agent.
Structured responses allow downstream tools to augment the results and enable the frontend to render purpose-built UI components.

\textbf{Clarify Variable}
Datasets can contain similar field names. For instance, the HuBMAP dataset~\cite{snyder_human_2019} includes \code{death\_event}, and \code{cause\_of\_death}, as well as \code{age\_value} and \code{gestational\_age\_value}.
If the user asks for a visualization of donor age and death, instead of inferring the intended field, the orchestrator can invoke the Clarify tool.
This tool will provide a free-text prompt requesting more information from the user, along with a structured list of candidate fields to select from.
Field descriptions are retrieved from the dataset schema and included in the prompt to ground the generated response.
The front-end system displays the message along with the discrete options they can select. 

\textbf{Explain}
The explanation tool provides the orchestrator with a flexible way to respond to the user. 
In addition to writing plain text, it can also embed data-driven values inline.
For instance, it can generate the response ``there are  \code{row\_count(donors)} donor records''. Before the tool call is finalized and sent to the frontend, the inline functions will be resolved with the underlying data. 
This strategy combines the flexibility of large language models with reliable, deterministic data queries.
On the front end, these structured data elements are rendered distinctly and include a tooltip to explain how the data was generated.

\textbf{Rebuff}
The various tools provided to the orchestrator agent in YAC cover a wide range of possible interactions. However, not every user request is possible within the system. Instead of generating an incorrect or unexpected response, the orchestration agent can invoke a rebuff tool, which informs the user that their request is not supported and suggests alternatives.
The system capabilities of YAC are provided to the orchestrator agent as context to help it determine when to invoke the rebuff tool. The same information is passed to the rebuff tool to help it generate suggested user queries, which are displayed in the chat interface as quick select buttons.

\subsection{Deployment and Technical Details}
YAC is free open-source software and has multiple instances deployed.
A standalone version is available (\url{https://hms-dbmi.github.io/udi-chat/}), requiring a user-provided API key.
Additionally, \textbf{YAC has been integrated into the HuBMAP data portal}, and is available to consortium members.
YAC is implemented as a web application.
The front-end (\url{https://github.com/hms-dbmi/udi-chat}) utilizes Vue and Quasar, as well as our visualization toolkit (\url{https://github.com/hms-dbmi/udi-grammar}).
The back-end handles agent orchestration, which is exposed through a Python API (\url{https://github.com/hms-dbmi/UDIAgent}).

\begin{figure}[tbh!]
	\centering
	\includegraphics[width=0.9\linewidth]{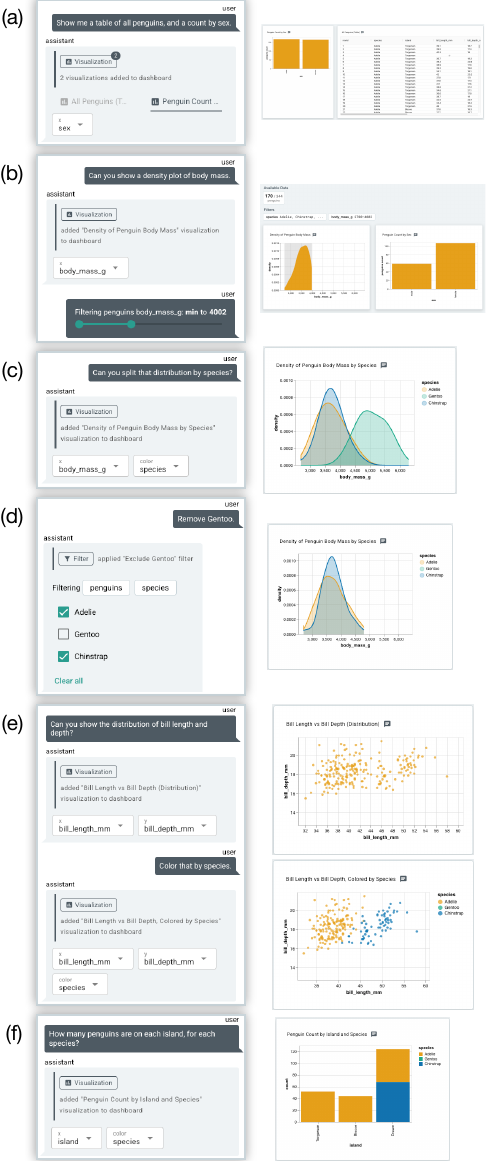}
	\vspace{-2mm}
	\caption{A hypothetical usage scenario of YAC.}
	\vspace{-8mm}
	\label{fig:case_study_penguins}
\end{figure}

\section{Usage Scenarios}

We illustrate YAC's capabilities with the familiar Palmer Penguins dataset \cite{gorman_ecological_2014, horst_palmerpenguins_2020}, where, instead of identifying a set of biological donors who meet various criteria we are identifying a set of penguins who meet various criteria.
The researchers are looking for a set of penguins from a single species, with an equal number of males and females, a smaller size, and are easy access.

They start by requesting \userprompt{Show me a table of all penguins, and a count by sex} and receive a tabular view plus a bar chart showing an equal number of male and female penguins (Figure~\ref{fig:case_study_penguins}a). The researcher then requests \userprompt{Can you show a density plot of body mass.}, and selects a lower range of values in the chart. This adds a filter widget to the conversation and updates the bar chart, which reveals that there are roughly half as many male penguins as female penguins with a body mass below 4000 grams (Figure~\ref{fig:case_study_penguins}b).

They remove the filter and request \userprompt{Can you split that distribution by species?} The new visualization shows that two penguin species (Chinstrap and Adelie) have a similar body mass distribution, while the third species (Gentoo) is larger (Figure~\ref{fig:case_study_penguins}c). Since the user wants to investigate smaller penguins, they request \userprompt{Remove Gentoo.} This results in a filter widget being added to the conversation that filters out penguins with a species of \textit{Gentoo} (Figure~\ref{fig:case_study_penguins}d).

After the user requests \userprompt{Can you show the distribution of bill length and depth?}, they can see in the new scatterplot that there appear to be two clusters of bill sizes . They are curious if this is due to the penguin species, so they ask YAC to \userprompt{Color that by species.} The new scatterplot indicates that each cluster belongs to a different species (Figure~\ref{fig:case_study_penguins}e).

The researcher has decided they will limit the species to either \textit{Adelie} or \textit{Chinstrap}, but have not decided which one. They decide to ask \userprompt{How many penguins are on each island, for each species?} This introduces a new bar chart that reveals that \textit{Adelie} penguins are on three different islands, but \textit{Chinstrap} is only found on one (Figure~\ref{fig:case_study_penguins}f).
They limit species to Adelie since they exist on all islands. Since all criteria are met, they download the list of penguins.

\section{User Study}

We conducted an expert evaluation of YAC with semi-structured interviews with task-based components.
Our study was approved by Harvard's IRB (No. IRB26-0088).
The 12 experts have experience with biomedical data portals. We conducted a thematic analysis across interviews, identifying recurring patterns and grouping them into themes. We also report summary statistics derived from conversation logs; the full logs are available in the supplementary material.

\textbf{Recruitment}
The 12 participants covered a variety of roles within the domain of biomedical research: research associate, postdoc, software developer, UX designer, professor, and scientific director.
Participants were selected based on one of two criteria: either (a) experience as computational biologists or bioinformaticians working with biological datasets, or (b) direct experience with biomedical data portals as software developers or UX designers.
We recruited participants through NIH consortia contacts and the Department of Biomedical Informatics at Harvard Medical School.
Recruitment and interviewing continued until thematic saturation was reached.

\textbf{Interview Process}
Our interview process was designed to elicit insight and feedback from domain experts. This included asking our participants to complete specific tasks in YAC; however, our goal was not to measure task completion rates, but rather to use these tasks as a probe to invite qualitative feedback from participants.
The interviews followed the same structure, with one interviewer leading the conversation and another primarily observing. 
Each interview began with the participants providing verbal informed consent. Then a brief discussion of the experts' role, background, and experience working with biomedical datasets. We then prompted users to interact with YAC in several ways: exploring system capabilities, understanding available datasets, constructing visualizations about donor demographics and death, filtering data based on specific criteria, and updating filters.
As participants interacted with the system we would ask contextual follow-up questions, such as: ``What do you expect to get as a result?'',  ``Is that what you expected to happen?'', ``What did you want instead?'', ``Why do you think the system is showing that?''. Finally, after going through our planned set of prompted tasks we invited the expert to explore the system however they wanted and solicited open-ended feedback.

\textbf{Qualitative Analysis}
The study results were analyzed by the two authors who conducted the interviews. Together they completed a thematic analysis \cite{braun_thematic_2019} of the interviews, which resulted in 13 interaction patterns and 5 themes, summarized in table \ref{tab:themes}. First, the two authors reviewed interview transcripts and videos and identified recurring interaction patterns. One author reviewed ten interviews, the other reviewed three. One interview was reviewed by both authors to establish coding consistency.
Next, they reconciled and merged overlapping patterns.
Finally, one author grouped interaction patterns into broader themes, and the other reviewed the groups. They discussed and refined the groupings iteratively until reaching a consensus.

\subsection{Results}

\begin{table*}[htbp]
\vspace{-2mm}
\caption{Themes and Interaction Patterns observed across participants.}
\vspace{-2mm}
\label{tab:themes}
\small
\begin{tabularx}{\textwidth}{p{3cm} X r}
\toprule
\textbf{Theme} & \textbf{Interaction Patterns} & \textbf{Participants} \\
\midrule

\multirow{2}{*}{Interpreted Output}
  & User is confused by the persistent filters that apply to all visualizations.
  & P1, P3, P5, P7, P8, P9, P12 \\
  & User understands system of persistent filters.
  & P1, P3, P5, P6, P7, P10 \\
\midrule

\multirow{3}{*}{Expected Output}
  & User requests new independent visualizations.
  & P1, P2, P3, P4, P6, P7 \\
  & User expects data filter to be embedded within each visualization.
  & P1, P4, P5, P7, P8, P9, P10, P12 \\
  & User requests specific visualization in response to general question.
  & P1, P2, P4, P5, P9 \\
\midrule

\multirow{3}{*}{\parbox{3cm}{Data, Trust, \& Transparency}}
  & User wants overview of available data and structure.
  & P1, P2, P4, P5, P6, P8, P9, P10, P12 \\
  & User wants additional information about data on interaction.
  & P1, P2, P4, P6, P7, P8, P10 \\
  & User does not trust the output of the system.
  & P4, P5, P6, P12 \\
\midrule

\multirow{2}{*}{\parbox{3cm}{Text Response Format}}
  & User states that the LLM response is too verbose and difficult to scan.
  & P1, P2, P3, P4, P5, P6, P7, P9, P10, P11 \\
  & User states that listing every field is excessive.
  & P1, P3, P4, P7 \\
\midrule

\multirow{3}{*}{System Boundaries}
  & User requests data, or asks how to download data.
  & P2, P6, P8, P10, P12 \\
  & User asks what YAC stands for.
  & P1 \\
  & User asks to start over, or return to a previous state.
  & P7, P8, P11 \\
\bottomrule
\end{tabularx}
\vspace{-5mm}
\end{table*}

Participants successfully completed a wide range of interactions with YAC.
Visualizations often displayed the requested information, users enjoyed the ability to filter data from the adjustment widgets, and they appreciated when the system told them their request was not possible or asked for clarification of variables with additional context.
However, our analysis focuses on moments of misalignment between user expectations and system behavior, as these yield the most actionable insights for future development.
These findings directly inform YAC's ongoing development as it is integrated into the HuBMAP data portal.

\textbf{Expected Output}
\textit{What output did the user expect when they formulated and sent their message?}
Many participants requested single independent visualizations as the output of their requests --- they appeared to have a mental model of how to interact with a chat-based interface before working with YAC.
In particular, filters were seen as something that was embedded within each visualization independently, not something applied to all visualizations globally. For instance, P10 had already created a bar chart that showed a breakdown of causes of death. They then requested \userprompt{Can you show me the causes of death for donors whose age range is between 25 and 50?}, which resulted in a new filter being applied to the existing visualization. However, this surprised the participant:

\pquote{P10}{Oh, I did not expect that. [Interviewer: What would you hope to see?] Another bar chart, similar to the first one, with a reduced count.}

This mental model was reflected in the way users specified their requests. For instance, in the scenario above, P10 reformulated their full  request, instead of saying ``filter that between age 25 and 50.''
Participants also requested modifications of visual encodings, for instance, to swap the axes. Some users desired a new visualization with the original intact while others wanted the visualization to be updated.
The expectation that the system would return a single visualization also influenced how they specified their queries. For instance, when given a broad question by the interviewer, such as ``summarize information about donor demographics and death'', many participants opted to request a single visualization.

Existing generic chat-based interfaces, where the output after each request is a new artifact, unlinked from previous constructions could explain this mental model. For instance, P7 suggested a UI change that reflects this expectation:

\pquote{P7}{Instead of having the chat on the left-hand side and then the result on the right-hand side, I would like to have it at the bottom. So after my chat, if I ask for visualizations, then visualizations will appear at the bottom.}





\textbf{Interpreted Output}
\textit{How did users interpret the output of the system?}
This theme is the counterpart to the previous theme.
In some cases, the output differed from what users anticipated, yet they were satisfied with the result, for instance, when the system rebuff tool indicated that the request was not possible.
A key finding is that users' expectations of independent static visualizations influenced their interpretation of the linked interactive visualizations.

Many users ran into a situation where the global cumulative filters resulted in a confusing application state. This became most apparent after multiple filters had been requested, eventually resulting in a combination where no data matched, and the visualizations were all empty. P5 ran into this situation after first applying a filter to remove female donors, then applying a new filter:

\pquote{P5}{[Interviewer: Does this make sense?] No it doesn't, because there are still zero females. This is not a great success story for Mr. Chatbot here. [\ldots] I guess if I remove this filter it will do better. No, it's still not much better. What I would like to have here is this all seeing, all dancing AI would provide me with some suggestions to increase the number of donors. Because clearly getting zero is a disappointment to most people.}

However, not all users struggled to adapt to how filters are applied.
Some were able to quickly understand how the linked filtering worked, while others took more time and prompting.
The interviewers did not tell participants how filters work, but did ask questions such as ``why do you think no data is showing?'', which likely helped participants review the interface more closely and notice the filter bar.
Once they understood the global filtering behavior, some participants came to prefer it
For instance, P10, who expected filters to be associated with each visualization independently, ultimately liked the ability to modify the filter of many visualizations on a single screen:

\pquote{P10}{[Interviewer: What do you think of the responses you're seeing now versus what you'd prefer?] I like this better actually. Given the space on my screen, I don't want to scroll up and down. I want to see a maximal number of plots without having to go left and right or up and down, I find that very annoying. I think [the filter widget] allowed me to sort of see the whole thing, and then I can filter further on here, which is really helpful. I think this is awesome.}

\textbf{Data, Trust, and Transparency}
\textit{Could users understand the data, and did they trust it?}
Users expressed a desire for additional information about the structure and overview of the data available. The structured text responses generally provided a sufficient overview in many situations, and users often referred back to the field names listed in the conversation text. In addition to descriptions of available field names, users wanted more information about the structure or relationships between tables. Users requested additional information about the definitions of the data. Although YAC provided them with field descriptions based on the dataset schema, they also desired information about the meaning of individual field values.
One user noticed the tooltip in the linked data element in text. However, they were expecting additional information about the data, not a description of how the data was generated. 

Some participants expressed a lack of trust in the system's accuracy. In some situations, this was due to the LLM making a mistake that the user noticed. For instance, P4 requested a visualization of a subset of data, and YAC generated the correct visualization, but did not apply a filter. This was their response after they requested in chat to \userprompt{show me slideseq datasets by organ}:

\pquote{P4}{So I was hoping to see something like this, but I am pretty sure that this is factually incorrect. [\ldots] I don't think that we have slide-seq data from all of these organs and in these quantities.  [\ldots] So what happened there is it just seemed to have ignored the slide-seq search term, where I was hoping that would apply a filter to find a particular kind of variant of the RNA-seq assay.}


In this case, the participant correctly identified the error. Once a participant ran into an issue, that lack of trust carried forward and led to a misinterpretation of correct data. In one instance, a participant assumed that sparse, but correct data was actually a filtering error. Users also wanted a way to verify that the system's output was correct. For instance P12 reflected:

\pquote{P12}{I think it would be helpful to have some touch points on truth where I can confirm that, hey, this is the way it really is, and it's not, you know, the AI doing something funny.}







\textbf{Text Response Format}
\textit{How did users respond to the format and length of system text responses?}
Many users found the text responses too verbose, sometimes making it difficult to pull out key information.
This happened primarily with the results generated by an LLM via the text explanation tool.
This tool includes instructions to the LLM to keep responses concise; however, this was not sufficient.
It is possible that formulating these instructions in another way could results in more desirable outputs.
It is difficult to know the best formulation in advance, and in practice this requires trial and error experimentation.

Another situation where the system response was too verbose was when the structured text response generated the field names for a table. In the case of HuBMAP this resulted in as many as 230 fields.
Since this part of the response is a programmatically generated we have more control over how we address this issue. The user interface could truncate the field list after a threshold, or display them within a searchable and collapsible user interface component.




\textbf{System Boundaries}
\textit{The user requests information, or to perform an action that isn't within YAC's capabilities.}
Although the large language model can respond to queries flexibly, it cannot perform arbitrary actions in this context. Users sometimes hit scenarios where they reached a boundary condition in the system --- they requested something that fell outside the system's current capabilities.

The simplest example of this is the request for an explanation of why the system is named YAC. The name appeared in the top-left of the user interface but did not include any description. This information was also not provided as context to the large language model, so when the user asked what YAC stood for, the LLM did not respond correctly.
A more common boundary that was hit relates to the data discovery task. Once datasets have been identified, the next step is to download the datasets for further analysis.
YAC includes an explicit download data button; however, many users still asked in chat either how to download data, or requested that the data be downloaded directly.
Again, YAC has not been configured to download data, or with the context of the user interface options, so the responses were not aligned well with system capabilities, and instead gave general instructions for downloading data on the HuBMAP data portal.

The last boundary condition that was reached was requests to revert the conversation state to a previous state, or in the extreme, completely reset the screen. Users could clear the screen by removing visualizations and filters one by one or by selecting the reset button, but they tended to request this via text first.

\section{Reflection}
We performed an analysis on the different technical dimensions of YAC with TDoPS \cite{jakubovic_technical_2023}. Although TDoPS was initially designed for programming systems, it has been applied to interactive user interfaces \cite{cutler_revisit_2026, mcnutt_mixing_2025}. We reuse McNutt et al.'s descriptions for dimensions \cite{mcnutt_mixing_2025}.
These dimensions provide a structured way to carefully consider the design choices we made and reflect on how our observations can be generalized to other systems.

\subsection{Interaction}
\textit{Which loops in the system are overlapping and how far apart are the corresponding gulfs of evaluation?}
There are two high-level categories for how users can interact with YAC. They can type text freely into the chat interface or interact with UI components and visualizations. The natural language interactions have a longer \textit{gulf of evaluation} --- it takes more time for the user to type a message and for the system to respond.
Alternatively, making changes with the adjustment widgets is much faster. Changing the visualization field or filter range results in much quicker updates to the visualizations in the dashboard. We designed these two types of interactions to be complementary. The expressivity of natural language allows for an easier entry point with flexible guidance towards the next step, and the UI components allow faster, more fine-grained course correction as the user learns more about the data.

The two modes of interaction share some overlap, but are still distinct. In the current version of YAC, users can only create visualizations from natural language requests in the chat. Data filters can be created either from the chat or from existing visualizations. Modifications of filters and visualizations can only occur through the user interface interactions. Requests to modify a visualization or filter from the chat will result in a new, modified visualization or filter being added to the current state. Finally, downloading the currently selected data is only possible through the user interface.
While building YAC, we faced a design tension on when to consider full parity of the system by supporting functionality through the chatbot and the user interface.
We intentionally designed YAC to have complementary interactions (\ref{r:complement}) between natural language requests and user interaction requests. However, our user study results indicate a few edge cases that will be improved in YAC. Additional tools will be added to give the orchestrator greater capabilities to perform the actions users requested. In addition, a self-explanation tool with documentation for all functionality in YAC could serve as an on-demand help tool when users ask how to perform tasks in the interface manually.

\subsection{Notation}
\textit{What notations are present and how do they interrelate?}
The two notations of the system mirror the two interaction modalities --- free text and UI selections. These two types of notations are connected by the underlying system of YAC. The agents that interpret the free text generate structured output. That structured output is interpreted by the system and presented to the user as a widget or visualization. The user can use these elements to make selections and modify the underlying data structure. In general, we present the same information about the application state to both the user and the system in the modality that is easiest for each person or agent to understand. The LLMs and YAC system can operate on structured representations of filters and visualizations, and humans can visually see when visualizations and filters are created. The way that these structured representations are presented to the user can vary depending on the context. For instance, visualization specifications are rendered as visualizations in the dashboard. However, those same specifications are rendered as adjustment widgets in the chat interface.

In our current design, natural language serves as the primary notation that drives the creation of the visualization dashboard, and user interface interactions enable refinement and exploration of the current dashboard.
One interesting difference between YAC's user interface interactions and natural language interactions is their stability with respect to input and output.
In other words, does similar input result in similar output?
Adjusting the filter slider by a small amount will change its appearance and the resulting data by a small amount.
However, for natural language, evaluating this property gets murkier.
Sometimes, large language models handle minor adjustments to input text well. 
For instance, sentences with typos can often still be interpreted correctly. Other times, seemingly insignificant changes to the input can result in a different output.
Furthermore, models are typically non-deterministic --- the same input will sometimes result in different outputs.
In YAC, we attempt to increase the stability of the output with the tool-based approach, which provides increased guardrails on the output.

Filtering actions have a unique aspect to them, which is that they can originate from both natural language requests and user selections in visualizations. Even though the source of the filters diverge, the notations we use to represent them are very similar --- both cases will result in a filter widget added to the chat interface. This uniform representation of notation was an intentional design choice to illustrate that the actions the agent takes to filter data have the same effect as the actions users can take to filter data.

\subsection{Conceptual Structure}
\textit{What is the shape of the notations at play and how do they relate to user goals?}
The behavior of YAC can be decomposed into primitive actions and the current application state. The application state can be modified through sequences of actions. The actions that YAC currently supports are chart creation, chart adjustment, filter creation, filter modification, and data download. The agent and user progressively take these actions to construct application states unique to the user's particular task. Multiple visualizations are composed to create a linked visualization dashboard. Multiple filters combine to define a more specific subset of data. Once a user has created a combination of filtering logic and verified that it satisfies the goals of the session, they can download the filtered data as the final step.

These actions can be initiated from both free text input and user interface interactions. One advantage of the free text input is that it offers flexibility in chart construction. Users can ask high-level questions or patterns they want to see in the data
\userprompt{Is there a relationship between donor height and weight?},
or request specific chart types \userprompt{Scatterplot donor height weight}. These higher-level prompts offer convenience to the user; they can keep their prompts simpler and let the agent respond with a reasonable response. However, since free text is so flexible, they can also specify elements of their chart creation in a detailed manner
\userprompt{entity: donors, mark: point, x-position: weight, y-position: height}.

\subsection{Customizability}
\textit{How can programs be modified?}
Users construct YAC's application state through sequences of free text queries and UI actions.
This sequence of actions constructs a custom dashboard that can be further interacted with. The current version of YAC focuses on this initial construction and exploration process. However, since we are storing these action sequences in a structured format, the tool is well-positioned to support further modifications. For instance, the visualization dashboard could be shared as a standalone chat-free interface that users can browse and filter. Conversation state could also be shared with a different user, allowing them to continue the conversation, or modify previous actions.

However, there are limits to the customizability of the visualization dashboard. YAC, intentionally, does not expose the full visualization grammar to the user. The adjustment widgets, as the name implies, allow minor adjustments of the visualization. They do not support arbitrarily changing visual encodings or data transformations. The filtering logic in YAC also currently imposes constraints. Multiple filters are assumed to be combined to create an intersection. E.g., find users who are male \textbf{and} have an age over 60. However, if you instead want to find all donors who are male \textbf{or} have an age over 60. This is not currently possible. This level of internal logic is not exposed to the user. However, we can still expand the capabilities of the underlying system with additional features without fully exposing complex technical details to the user. 

\subsection{Complexity}
\textit{How is complexity dealt with through design and automation?}
A grand challenge in designing technical systems is to support complex interactions while retaining a simple and intuitive interface. In YAC, we explore the potential of natural language for addressing this challenge. Our current system is capable of constructing linked, multi-view visualizations from multiple different data tables with targeted adjustment widgets --- all from a few natural language questions.
However, one challenge with this approach that may not be immediately obvious is that even if something can be expressed as a natural language request, it does not mean that the system is capable of meeting that request.
For instance, YAC currently assumes that filters combine with intersection logic. This is an invisible boundary that is not immediately apparent in the interface.
An earlier version of YAC did not include text-based responses; hitting the boundary of the system was more jarring.
Since adding the structured text-based responses, especially the rebuff tool which was designed specifically for this purpose --- the problem has been ameliorated.
Still, the underlying capabilities of YAC can be further enhanced to support additional functionality.
Designing a successful system still requires understanding ways that users will try to use the tool and, consequently, the functionality that the underlying system must support.


\subsection{Errors}
\textit{What are they and how are they handled? }
There are three types of errors that we consider in YAC. All the errors relate to the free-text input and structured response of the agent. The first type of error is a syntactic mistake in the output where the structured output is invalid. These errors can be prevented altogether by passing a desired structure (e.g., with a JSON Schema), which guarantees that the LLM will only generate responses that adhere to that schema.
The second broad category of errors is syntactically valid outputs that are semantically incorrect. E.g., the agent responded with a visualization specification for a scatterplot, but that scatterplot did not answer the question the user had. In YAC, we aim to minimize these errors by utilizing our fine-tuned model and providing detailed information about the data fields to the multi-agent system. However, as a fallback, we provide the ability for the user to adjust the target data fields in visualizations and filters. 
The final category involves resolving ambiguity in the user prompts. This category is not an error in the traditional sense, but it fits into this theme because it represents a gap between the generated response and the desired response. For instance, ambiguous terms could be used to request data to be filtered \userprompt{filter to old donors}. In this scenario, there is no single correct answer. \textit{Old} will mean different things to different users of the system. 
In this example, an inherently ambiguous term was used, even if the user had a precise number in mind. However, the reverse can also occur. The user requests a precise filter \userprompt{filter to donors older than 85}, but does not know the actual boundary they want. They need to start somewhere, and once the filter is applied, they can see the effect on their visualization dashboard. They may realize that with this filter applied, they have too few records selected and need to expand it.
YAC includes support before and after generating a result. The variable clarification response can preemptively request that the user provide additional information to disambiguate their requests.
After the results has been generated adjustment widgets provide additional refinement.
The agent can apply a filter range, and the user can quickly make adjustments to that without typing and sending a new free-text request.

\subsection{Adoptability}
\textit{What socio-technical (e.g., learnability) dimensions are considered?}
One of the primary goals in designing YAC is to create an interface with a lower barrier of entry, in other words, one with a gentler learning curve.
This goal motivated our choice to use natural language text as the primary input for YAC. We assume that users of YAC are expert researchers and have an initial question. Thus, they will be able to write that question down. The second strategy to help with the learnability of YAC is to progressively build up the complexity of the interface. When the session begins, there will be no visualizations and no filters applied. As the user asks questions, filters, and visualizations will be progressively added to the interface. This approach introduces user interface complexity --- just in time --- right when the user needs it, as opposed to traditional interfaces, which display all features upfront.

Several other technical and economic factors also influence the tool's adoptability. The input and output of the tool are designed to be simple, easy to work with, and utilize existing standards. The input to YAC is a data package that defines the existing entities and their relationships. We use an existing standard, instead of defining our own standard, to utilize existing tools and datasets.
The next consideration is how YAC can be accessed. YAC is an open-source tool developed as a front-end application (\url{https://hms-dbmi.github.io/udi-chat/}) and can be accessed by anyone with a personal computer and internet access.
However, actually using the tool incurs an economic cost, due to the large language models
YAC's multi-agent system currently makes calls to OpenAI's API service, which charges based on token usage. This has a potential impact on adoptability, because it means that users must either provide their own API key and pay OpenAI, which is the case for our standalone version. Alternatively a central organization has to pay OpenAI for broader use of YAC, which is the case for the deployment integrated into the HuBMAP data portal.

\section{Discussion}

There are two interesting technical decisions we made while designing YAC that could apply to other systems that incorporate generative AI with visualization interfaces.

\subsection{Fine-Tuning versus Tool-Based Generation}
An earlier version of YAC used a model we fine-tuned with the DQVis Dataset \cite{lange_dqvis_2025} to generate visualization specifications.
This model had some success; however, iterating on the model required time and computational resources. Any adjustment required updating the training data, and retraining the LLM, which took days to weeks.
Alternatively, the tool-based generation of visualizations could be modified within minutes by adjusting the available tools and system instructions.

\subsection{Generating Structured Output}
There are several possible strategies for LLMs to generate visualizations at different levels of abstraction.
For instance, it can generate code (e.g., Python or JavaScript) that must be executed to create the visualization. Alternatively, it can generate a specification (e.g., JSON) that provides all the necessary information to create the visualization. For our project, we decided to target specifications in the form of a grammar we designed. We believe there are several advantages to this choice.
The first advantage is the guardrails that this approach provides. The visualization agent is designed to create only visualization specifications.
The interface will ignore anything generated outside the bounds of our JSON Schema.
The second advantage is that the structured nature of the specification outputs facilitates the integration of responses into a larger system.
Our multi-agent system generates structured outputs that define visualizations and filters. The well-defined structure of these outputs allows the system to complete several integration tasks, tying multiple outputs together. Adjustment widgets can display and update values in filters. Visualizations can be rendered, but also modified with adjustment widgets. Finally, linked filtering between visualizations and filter widgets is possible by updating the visualization specifications with simple rule-based logic.

\section{Limitations and Future Work}
The current version of YAC has several technical limitations that could be addressed in the future. 

\textbf{Multi-Entity Relationships}
YAC currently supports linked filtering across entities. However, it only includes logic for tables that are directly related. It does not handle tables that are linked through one or more intermediate tables. However, such capability could be integrated into a future version of YAC.

\textbf{Scalability Performance}
YAC currently only supports relatively small data files, on the order of tens of thousands at most. Since it is designed for analysing biomedical data repository metadata, not the raw data, this limitation is not an immediate constraint. However, we can improve the scalability of the data transformation by utilizing modern technology such as DuckDB\footnote{https://duckdb.org/} or Mosaic \cite{heer_mosaic_2024}.

\textbf{Filter Intersection}
The filtering logic currently assumes that filters always combine to create a single subset of data defined by the intersection of all filters. We could expand the capabilities of YAC to support more flexible filtering logic, such as unions of filters, and the creation of multiple parallel selections.

\textbf{Beyond Metadata}
YAC was intentionally designed to support metadata tables commonly found in biomedical data repositories. 
YAC could be further enhanced to analyze genomics and single-cell data, potentially enabling new ways to integrate data discovery with data analysis.

\section{Conclusion}

In conclusion, we developed  YAC, a prototype system that integrates elements of a traditional data discovery interface and visualizations with a chat-based system.
YAC offers a new paradigm that allows researchers to flexibly navigate complex, multi-entity biomedical metadata without being constrained by predefined interfaces.
An effective universal discovery interface (UDI) such as YAC that enables natural language queries for data discovery also has bigger implications for the design and implementation of future biomedical data resources.
First, a UDI could be deployed as an initial lightweight discovery interface for new data resources, accelerating the value gained from them.
Second, a natural language interface is a formidable method to collect information about user needs and insights --- the design inherently requires users to provide their questions to the system.
We envision that future data resources will be launched with UDIs and information gathered will be used to build, either manually or with the help of AI, custom interfaces for common data discovery needs.

\acknowledgments{

This work was in part supported by ARPA-H award AY2AX000028. The authors wish to thank the volunteers who participated in the user study for their feedback and the members of the HIDIVE Lab for their support throughout this project. 
}

\bibliographystyle{abbrv-doi-hyperref}
\bibliography{ref, manual_refs}

\appendix 
\crefalias{section}{appendix} 

\end{document}